\documentclass[12pt]{article}

\usepackage{amsmath,amssymb}
\usepackage{amsthm}

\theoremstyle{definition}\newtheorem{defin}{Definition}[section]
\theoremstyle{plain}\newtheorem{theo}[defin]{Theorem} 
\theoremstyle{plain}\newtheorem{prop}[defin]{Proposition} 
\theoremstyle{plain}\newtheorem{lem}[defin]{Lemma} 
\theoremstyle{definition}\newtheorem{ex}[defin]{Example}

\title{Fermat Principle in Finsler Spacetimes}
\author{Volker Perlick\thanks{TU Berlin, Sekr. PN 7-1, 10623 Berlin, Germany.
Email: vper0433@itp.physik.tu-berlin.de}}
\date{}

\begin{document}

\maketitle


\begin{abstract}
	It is shown that, on a manifold with a Finsler metric of Lorentzian
	signature, the lightlike geodesics satisfy the following variational
	principle. Among all lightlike curves from a point $q$ (emission event)
	to a timelike curve $\gamma$ (worldline of receiver), the lightlike 
	geodesics make the arrival time stationary. Here ``arrival time'' refers 
	to a parametrization of the timelike curve $\gamma$. This variational
	principle can be applied (i) to the vacuum light rays in an alternative
	spacetime theory, based on Finsler geometry, and (ii) to light rays in
	an anisotropic non-dispersive medium with a general-relativistic spacetime
	as background. 
\end{abstract}


\section{Introduction}\label{sec:intro}
The versions of Fermat's principle that can be found in standard text-books
refer to stationary situations, both in general relativity (see e.g. Landau and
Lifshitz \cite{LL62}) and in ordinary optics (see e.g. Kline and Kay \cite{KK65}). 
The goal is to determine the path of a light ray from one point in space to another 
point in space, under the influence of a time-independent gravitational field or a 
time-independent optical medium. A basic idea of how to generalize these standard 
versions of Fermat's principle to non-stationary situations is due to Kovner 
\cite{Ko90}. He considered an arbitrary spacetime in the sense of general 
relativity, i.e., a manifold with a pseudo-Riemannian metric of Lorentzian 
signature that need not be stationary. He fixed a point (emission 
event) and a timelike curve (worldline of receiver) in this spacetime. The variational 
principle is to find, among all lightlike curves from the point to the timelike 
curve, those which make the arrival time stationary. Here ``arrival time'' refers 
to an arbitrary parametrization of the timelike curve. It was proven in \cite{Pe90} 
that the solution curves of this variational principle are, indeed, precisely 
the lightlike geodesics. Kovner's variational principle can be viewed as a 
general-relativistic Fermat principle for light rays that are influenced by 
an arbitrarily time-dependent gravitational field with no optical medium. 
For some applications one has to use the time-reversed version of this 
variational principle, which is mathematically completely equivalent. Under 
time reversion an emission event turns into a reception event, and the worldline 
of a receiver turns into the worldline of an emitter, so each solution to the 
variational principle corresponds to an image of the emitter that is seen at 
the chosen reception event. In this time-reversed version, Kovner's 
variational principle can be used for investigating gravitational 
lensing situations, see Kovner's original work and, e.g., \cite{SEF92, Pe00, Pe04}.

Without mathematical modifications, Kovner's version of Fermat's principle 
also applies to the case that, in addition to the (time-depedent) gravitational 
field, there is a (time-dependent) isotropic non-dispersive optical medium. 
Such a medium can be characterized by an index of refraction that depends 
on the spacetime-point but neither on spatial direction nor on frequency. It 
was observed already in 1923 by Gordon \cite{Go23} that the light rays in 
such a medium are the lightlike geodesics of a pseudo-Riemannian metric of 
Lorentzian signature which is called the \emph{optical metric}. So all one 
has to do in order to apply Kovner's variational principle to this situation 
is to replace the spacetime metric with the optical metric. 

In this paper it is our goal to generalize Kovner's version of Fermat's principle
to the case that the light rays are the lightlike geodesics of a Finsler metric,
rather than of a pseudo-Riemannian metric, of Lorentzian signature. There are
two physical motivations for such a generalization.

First, the theory of Finsler metrics is often considered as an alternative spacetime 
theory which modifies general relativity by allowing for the possibility that the 
vacuum is spatially anisotropic, even in infinitesimally small neighborhoods. Although
up to now there is no observational evidence in this direction, such a modification 
of general relativity has found a lot of interest among theorists. An 
extensive list of the pre-1985 literature can be found in Asanov's book on
the subject \cite{As85}. The variational principle to be established
in this paper applies to light rays in such a modified spacetime 
theory, without a medium, and may be used, e.g., as a tool for investigating 
(hypothetical) Finsler gravitational lenses. 

Second, Finsler metrics naturally appear in optics of anisotropic non-dispersive
media. More precisely, if one performs, on a spacetime in the sense of general
relativity, the passage from Maxwell's equations to ray optics in a medium characterized
by a dielectricity tensor and a permeability tensor that have to satisfy some regularity
conditions, one finds the following, see \cite{Pe00}, Section 2.5. Corresponding to the 
fact that such an anisotropic medium is birefringent, there are two different Hamiltonians 
for the light rays. Each of the two Hamiltonians is homogeneous of degree two with respect 
to the momenta. Providing the validity of regularity conditions, the second derivative 
of each Hamiltonian with respect to the momenta is non-degenerate and of Lorentzian
signature. This means that each of the two types of light rays can be characterized 
as the lightlike geodesics of a Finsler metric of Lorentzian signature. In other words,
the variational principle to be established in this paper applies to light
rays in an anisotropic non-dispersive medium that is in arbitrary motion on the 
background of a spacetime in the sense of general relativity. Various earlier versions
of Fermat's principle are contained as special cases. E.g. if the anisotropic 
non-dispersive optical medium is at rest in an inertial system, we recover the versions
of Kline and Kay \cite{KK65}, Section III.11, and Newcomb \cite{Ne83}; if it is moving 
with temporarily constant velocity with respect to an inertial system, we recover the 
version of Glinski{\v \i} \cite{Gl80}. Moreover, it should be noted that Fermat's 
principle can be applied not only to light rays but also to sound rays. E.g., there 
are versions of Fermat's principle for sound rays in an anisotropic elastic medium 
that is at rest in an inertial system (see Babich \cite{Ba61}, Epstein and {\' S}niatycki 
\cite{ES92} and e.g. {\v C}erven{\' y} \cite{Ce02}) and 
in a fluid flow that is moving with temporarily constant velocity with respect to an 
inertial system (see Ugin{\v c}ius \cite{Ug72}). Our variational principle indicates 
how these results can be generalized to the case of media in arbitrary motion on a 
general-relativistic spacetime.

There is one earlier version of Fermat's principle for time-dependent anisotropic
non-dispersive media by Godin and Voronovich \cite{GV04}. In contrast to the approach 
presented here, Godin and Voronovich make no use of Finsler geometry, and they
do not work in the setting of a spacetime manifold with unspecified topology; they
rather assume that spacetime is a product of 3-dimensional space and a time axis,
and they make strong use of this product structure. The relation between the 
formulation of Godin and Voronovich and the one presented here will be clarified 
in Section \ref{sec:stat} below.  

It should be mentioned that even more general versions of Fermat's principle,
allowing not only for time-dependence and anisotropy but also for dispersion,
are already available. In \cite{Pe00}, Section 7.3, a variational principle 
in the spirit of Kovner is established for rays that are determined by a Hamiltonian 
function on the cotangent bundle over spacetime that has to satisfy only a 
certain regularity condition. (A similar, though in various technical respects 
different, variational principle was suggested by Voronovich and Godin 
\cite{VG03}.) One could establish the variational principle for 
lightlike geodesics in Finsler spacetimes by demonstrating that it is a 
special case of the one given in \cite{Pe00}, Section 7.3. However, this
is technically more difficult (and less instructive) than giving the proof
directly. Therefore in this paper we choose the latter way.

The paper is organized as follows. In Section \ref{sec:assumptions} we fix
our notation. In Section \ref{sec:deffinsler} we specify 
our definition of a Finsler spacetime and collect some basic mathematical
facts that will be needed later. In Section \ref{sec:fermat} we formulate
Fermat's principle for lightlike geodesics in Finsler spacetimes as a 
mathematical theorem. Section \ref{sec:proof} is devoted to the proof of
this theorem. In Section \ref{sec:stat} we consider some special cases;
in particular we demonstrate that, under appropriate additional assumptions,
our version of Fermat's principle reduces to the one of Godin and Voronovich 
\cite{GV04}.


\section{Notations and conventions}\label{sec:assumptions}
We denote the tangent space to a manifold $M$ at the point $x$ by $T_x M$. A circle 
over the $T$ indicates that the zero vector is omitted, i.e. $\overset{o}{T} {}_x M 
:= T_xM \setminus \{ 0 \}$. The cotangent space at $x$ is denoted by $T^* _x M$. 
For the tangent bundle we write $TM$ and for the cotangent bundle we write $T^*M$, 
i.e., $TM := \cup _{x \in M} T_xM$ and $T^*M := \cup _{x \in M} T^*_xM$. 
Correspondingly we use the notation $\overset{o}{T} M := \cup _{x \in M} 
\overset{o}{T}{}_xM$.

We denote points in an $N$-dimensional manifold $M$ by $x$, points in $TM$ by $(x,v)$ 
and points in $T^*M$ by $(x,p)$. Here $x= (x^1, \dots , x^N )$ stands for a coordinate 
tuple, in an unspecified local chart, and $(x,v)=(x^1, \dots , x^N, v^1 , \dots , v^N)$ 
and $(x,p)=(x^1, \dots , x^N, p_1 , \dots , p_N)$ stand for coordinate tuples in
the induced natural charts. This identification of points with coordinate tuples
is non-puristic but notationally convenient (and quite common in the literature 
on Finsler structures or, more generally, on Lagrangian and Hamiltonian equations).
Correspondingly we use the familiar index notation for tensor fields, with 
Einstein's summation convention for latin indices running from 1 to $N$,
and occasionally for greek indices running from 1 to $N-1$. 

It is to be emphasized
that the use of coordinate notation does not mean any restriction as to the 
global topology. E.g., if we write an equality of vector fields, $A^i (s) = 
B^i (s)$, along a curve parametrized by $s$, we do not imply that the 
curve can be covered by a single chart; rather, we mean the invariant equation
that takes the given form in any chart which covers the point with parameter 
value $s$. 


\section{Definition of Finsler spacetimes}\label{sec:deffinsler}
Finsler geometry was originally introduced as a generalization of Riemannian geometry, 
i.e., for positive definite metrics. This theory of positive definite Finsler metrics
is detailed, e.g., in the text-book by Rund \cite{Ru59}. The systematic study of
indefinite Finsler metrics, in particular of Finsler metrics with Lorentzian signature, 
began with a series of papers in the 1970s by John Beem. In the first paper of this 
series \cite{Be70}, Beem defines indefinite Finsler structures in terms of a sufficiently
differentiable Lagrangian function $L : \overset{o}{T} M \rightarrow \mathbb{R}$ that is 
positively homogeneous of degree two and whose Hessian is non-degenerate (with the desired 
signature). This is an appropriate definition for the purpose of the present paper, 
so we will adopt it in the following. (For convenience we will require that $L$ is 
of class $C^{\infty}$, whereas Beem allows for a less restrictive differentiability
condition.) It should be emphasized that large parts of the physical literature on 
indefinite Finsler metrics is based on weaker notions. E.g., Asanov \cite{As85} 
defines Finsler structures in terms of a function F that is defined, as a sufficiently
differentiable real-valued function, only on an unspecified open subset of $\overset{o}{T}M$;
the elements of this subset are called the ``admissible vectors'' by Asanov. On 
this subset, $F$ is assumed to be strictly positive and positively homogeneous of degree 
one, and the Hessian of $F^2$ is assumed to be non-degenerate (with the desired
signature). Clearly, this definition is weaker than Beem's. If $L$ satisfies the 
assumptions of Beem, $F := \sqrt{\pm L}$ satisfies the assumptions of Asanov, 
with the admissible vectors given as the set on which $\pm L$ is positive. (One
has to choose the plus or the minus sign, depending on the choice of signature.)
The reason for us to use Beem's definition, rather than Asanov's, is the following. 
On the basis of Beem's definition we can define lightlike vectors as those vectors 
on which $L$ takes the value zero, and we can define lightlike geodesics as those 
solutions of the Euler-Lagrange equation whose tangent vectors are lightlike. On the 
basis of Asanov's definition, the notion of lightlike vectors cannot be defined 
(in an invariant way), because $F^2$ is non-zero on the admissible vectors and 
nothing is said about the extendability of $F^2$ beyond the admissible vectors. 
As a consequence, there is no (observer-independent) notion of light rays in 
Asanov's setting, so the question of whether light rays satisfy a variational 
principle cannot even be formulated. For this reason, Asanov's definition is 
too weak for the purpose of the present paper.

Therefore, we adopt as the basis for our discussion the following definition.

\begin{defin}\label{def:finsler}
	An $N$-dimensional \emph{Finsler spacetime} is a pair $(M,L)$ where
\begin{itemize}
\item[(a)]
	$M$ is an $N$-dimensional real second-countable and Hausdorff $C^{\infty}$
	manifold.
\item[(b)]
	$L: \overset{o}{T}M \longrightarrow \mathbb{R}$ is a $C^{\infty}$ function
	that satisfies the following conditions.
\begin{itemize}
\item[($\,$i$\,$)]
	$L(x, \cdot )$ is positively homogeneous of degree two, 
\begin{equation}\label{eq:homL}
	L(x,kv) \, = \, k^2 \, L(x,v) \quad \text{ for all } k \in \: ] 0, \infty [ \: ;
\end{equation}
\item[(ii)]
	the \emph{Finsler metric} 
\begin{equation}\label{eq:defg}
	g_{ij}(x,v) \, := \, \frac{\partial ^2 
	L(x,v)}{\partial v^i \partial v^j}
\end{equation}
	is non-degenerate and of Lorentzian signature $(+ , \dots , + , - ) \,$.
\end{itemize}
\end{itemize}
\end{defin}
We call $L$ the \emph{Finsler Lagrangian} henceforth. It is a standard exercise to
check that condition (b) of Definition \ref{def:finsler} implies the identities
\begin{gather}
\label{eq:hom1}
	\frac{\partial L(x,v)}{\partial v ^k} \, v ^k \; = \; 2 \, L(x,v) \; ,
\\[0.3cm]
\label{eq:hom2}
	\frac{\partial L(x,v)}{\partial v ^i} \; = \;  g_{ij}(x,v) \, v^j \; ,
\\[0.3cm]
\label{eq:hom3}
	L(x,v) \; = \; \frac{1}{2} \, g_{ij}(x,v) \, v^i \, v^j \; .
\end{gather}

We call a Finsler spacetime \emph{isotropic} at the point $x$ if the Finsler metric 
$g_{ij}(x,v)$ is independent of $v$. If this is true for all $x$, $g_{ij}$ is a
pseudo-Riemannian metric of Lorentzian signature, i.e., it can be interpreted, in
the case $\mathrm{dim} (M)=4$, as
the spacetime metric in the sense of general relativity (but also as the optical
metric in an isotropic medium on a general-relativistic background). In this case,
the set $\big\{ (x,v) \in \overset{o}{T} _x M \, \big| \, L(x,v)<0 \big\}$ has two 
connected components for every $x \in M$; similarly, the boundary of this set in 
$\overset{o}{T}_x M$ has two connected components, a ``future light cone'' and a 
``past light cone''. In an arbitrary Finsler spacetime, however, the set $\big\{ (x,v) 
\in \overset{o}{T} _x M \, \big| \, L(x,v)<0 \big\}$ may have arbitrarily many 
connected components; correspondingly, there may be arbitrarily many ``light cones''. 
Finsler spacetimes with two or more future light cones at each point are probably
not of physical interest. (Note that a birefringent medium is not described by
one Finsler structure with two future light cones, but rather by two Finsler
structures with one future light cone for each, see Example \ref{ex:hamilton} below.)
However, there is no mathematical reason to exclude them. For the formulation of 
Fermat's principle we will just have to select one light cone, and we will need 
the results stated in the following proposition. Recall that a subset $S$ 
of a vector space is called \emph{convex} if $kS+(1-k)S \subseteq S$ for all $k \in 
\:]0,1[ \:$ and that it is called a \emph{cone} if $kS \subseteq S$ for all $k \in 
\:]0,\infty[ \:$.

\begin{prop}\label{prop:cone}
Fix a point $x$ in a Finsler spacetime $(M,L)$. Let $Z_xM$ be a connected component
of the set $\big\{ (x,v) \in \overset{o}{T} _x M \, \big| \, L(x,v)<0 \big\}$ and let
$C_x M$ be the boundary of $Z_xM$ in $\overset{o}{T}_xM$. Then the following is true.
\begin{itemize}
\item[\emph{(a)}]
$Z_x M$ is an open convex cone in $T_xM \,$.
\item[\emph{(b)}]
$C_xM$ is a cone in $T_x M$ and a closed $C^{\infty}$ submanifold of codimension 
one in $\overset{o}{T}_xM \,$.
\item[\emph{(c)}]
{\Large $\frac{\partial L(x,w)}{\partial w^j}$}$\, u^j \, = \, g_{ij}(x,w)\, w^i \, u^j 
\, < \, 0 \,$ for all $(x,w) \in C_xM$ and $(x,u) \in Z_x M \,$.  
\end{itemize}
\end{prop}
\begin{proof}
To prove part (a), it suffices to prove that $Z_xM$ is convex because the rest
of the claim follows directly from the definition. Beem \cite{Be70} has shown
that, as a consequence of Definition \ref{def:finsler} (b), each connected component 
of the set $\{ \, (x,v) \in \overset{o}{T} _x M \, | \, L(x,v)<-1 \, \}$ is convex.
As $L(x,v)$ is homogeneous with respect to $v$, this obviously implies that each 
connected component of the set $\{ \, (x,v) \in \overset{o}{T} _x M \, | \, 
L(x,v)<-c^2 \, \}$ is convex, for every $c>0$. As any two points in $Z_xM$ are 
contained in such a component, for some $c>0$, this proves that $Z_x M$ is convex. --
To prove part (b), we first observe that, by (\ref{eq:hom2}), $\partial L (x,v) / 
\partial v^i$ has no zeros on $\overset{o}{T}_x M$, so the set $\{(x,v) \in 
\overset{o}{T}_xM \, | \, L(x,v) =0 \}$ is a $C^{\infty}$ submanifold of 
codimension one in $\overset{o}{T}_xM$. By definition, $C_xM$ is a subset of 
this submanifold. As $C_x M$ is the boundary of the open set $Z_x M$ in 
$\overset{o}{T}_xM$, it must be a connected component, or the union of 
several connected components, of this submanifold, hence it is a cone in 
$T_xM$ and a closed $C^{\infty}$ submanifold of $\overset{o}{T}_xM$. --
To prove part (c), we consider $(x,u) \in Z_x M$ and a sequence of vectors
$(x,w_J) \in Z _x M$ such that $(x,w_J) \to (x,w) \in C_xM$ for $J \to \infty$.
As $Z_xM$ is a convex cone, $(x,w_J+ku) \in Z_xM$, i.e.,
$L(x,w_J+ku)<0$, for all $k \in \: ]0, \infty [ \,$. Sending $J$ to infinity, we
find 
\begin{equation}\label{eq:wineq}
	L(x,w+ku) \, \le \, 0 \quad \text{ for all } k \in \: ]0, \infty [ \; .
\end{equation}
On the other hand, Taylor's theorem yields
\begin{equation}\label{eq:Taylor1}
	L(x,w+ku) \, = \, L(x,w) \, + \, \frac{\partial L(x,w)}{\partial w^i} \,
	k \, u^i \, + \, \frac{1}{2} \, \frac{\partial ^2 L (x,w)}{\partial w^i \partial w^j}\,
	k^2 \, u^i \, u^j \, + \, O \big( k^3 \big) \; ,
\end{equation}
which can be rewritten, with the help of (\ref{eq:defg}) and (\ref{eq:hom2}), as
\begin{equation}\label{eq:Taylor2}
	L(x,w+ku) \, = \, L(x,w) \, + \, k \, g_{ij}(x,w) \, w^j \, u^i \, + \, \frac{1}{2} \,
	k^2 \, g_{ij}(x,w) \, u^i \, u^j \, + \, O \big( k^3 \big) \; .
\end{equation}
As $L(x,w)=0$ by assumption, (\ref{eq:wineq}) and (\ref{eq:Taylor2}) imply
\begin{equation}\label{eq:lim}
	g_{ij}(x,w) \, w^j \, u^i \, + \, \frac{1}{2} \,
	k \, g_{ij}(x,w) \, u^i \, u^j \, + \, O \big( k^2 \big) \, \le \, 0
	\quad \text{ for all } k \in \: ]0, \infty [ \; .
\end{equation}
By considering arbitrarily small $k$, we find $g_{ij}(x,w) \, w^j \, u^i \, \le \, 0 \,$.
Assume that $g_{ij}(x,w) \, w^j \, u^i \, = \, 0 \,$, i.e., that $u$ is
perpendicular to the lightlike vector $w$ with respect to the Lorentzian metric
$g_{ij}(x,w)$. As our hypotheses exclude the case that $u$ is a multiple of $w$,
this assumption implies $g_{ij}(x,w) \, u^j \, u^i \, > \, 0 \,$ which contradicts
(\ref{eq:lim}). We have thus proven that $g_{ij}(x,w) \, w^j \, u^i \, < \, 0 \,$.
\end{proof}

The non-degeneracy of the Finsler metric guarantees that the Euler-Lagrange equation
\begin{equation}\label{eq:EL}
	\frac{d}{ds} \, 
	\frac{\partial L \big( \lambda (s), \dot{\lambda} (s) \big)}{
	\partial \dot{\lambda}{}^i (s)} 
	\, - \,
	\frac{\partial L \big( \lambda (s) , \dot{\lambda} (s) \big)}{
	\partial \lambda ^i (s)} 
	\, = \, 0
\end{equation}
has a unique solution $s \mapsto \lambda (s)$ to each initial condition 
$\big( \lambda (0),{\dot{\lambda}}(0) \big) \, = \, (x,v) \, \in \overset{o}{T}M$. 
The solutions to (\ref{eq:EL}) are called the \emph{affinely parametrized geodesics} 
of the Finsler spacetime $(M,L)$. The homogeneity of $L$ guarantees that for a 
geodesic the equation $L \big( \lambda (s) , \dot{\lambda} (s) \big) = 0$ holds 
for all $s$ if it holds for $s=0$. Geodesics with this property are called 
``lightlike''. When formulating Fermat's principle it is our goal to characterize 
lightlike geodesics by a variational principle. As the affine parameter along a 
lightlike geodesic has no particular physical significance, we may allow for 
arbitrary reparametrizations. Instead of (\ref{eq:EL}), we then get the equation 
for \emph{arbitrarily parametrized geodesics}
\begin{equation}\label{eq:ELpar}
	\frac{d}{ds} \, 
	\frac{\partial L \big( \lambda (s), \dot{\lambda} (s) \big)}{
	\partial \dot{\lambda}{}^i (s)} 
	\, - \, 
	\frac{\partial L \big( \lambda (s) , \dot{\lambda} \big)}{
	\partial \lambda ^i (s)}  
	\, = \, w(s) \, 
	\frac{\partial L \big( \lambda (s), \dot{\lambda} (s) \big)}{
	\partial \dot{\lambda}{}^i}
\end{equation}
where $w$ is an unspecified function of the curve parameter.

Moreover, the non-degeneracy of the Finsler metric implies that the 
equation  
\begin{equation}\label{eq:Legendre}
	p_i \, = \, g_{ij}(x,v) \, v^j 
\end{equation}
defines a map $\overset{o}{T}M \rightarrow \overset{o}{T}{}^* M , (x,v) \mapsto 
(x,p)$ that is locally invertible. If this map is even a global diffeomorphism, 
we get a globally well-defined Hamiltonian $H: \overset{o}{T}{}^* M \rightarrow 
\mathbb{R}$ by Legendre transforming $L$, 
\begin{equation}\label{eq:defH}
	H(x,p) \, = \, v^i \, \frac{\partial L (x,v)}{\partial v^i} 
	\, - \, L(x,v) \, = \, L(x,v) \; ,
\end{equation}
where $(x,p)$ and $(x,v)$ are related by (\ref{eq:Legendre}). $H$ is positively 
homogeneous of degree two, $H(x,kp)=k^2 H(x,p)$ for $k>0$, and its Hessian 
$\partial ^2 H(x,p) / (\partial p_i \partial p_j)$ is the inverse of $g_{ij}(x,v)$,
thus non-degenerate with Lorentzian signature. The projections to $M$ of the 
solutions of Hamilton's equations with $H=0$ are precisely the affinely parametrized
lightlike geodesics. So we may work in a Hamiltonian formalism on the
cotangent bundle, rather than in a Lagrangian formalism on the tangent
bundle, whenever we wish to do so.  
 
\begin{ex}\label{ex:hamilton}
The physically most relevant class of Finsler spacetimes is given by Lagrangians
of the form
\begin{equation}\label{eq:Lex}
	L(x,v) \, = \, \frac{1}{2} \, \big( \, \ell (x,v)^2 \, - \, 
	U_i (x) \, U_j (x) \, v^i \, v^j \, \big) 
\end{equation}
where 
\begin{itemize}
\item[(a)]
	$\ell (x, k v ) \, = \, k \, \ell (x,v)\:$ for $k \in \: ]0, \infty [ \:$,
\item[(b)]
	{\Large $\frac{\partial ^2 \ell(x,v)^2}{\partial v^i v^j}$}$\, w^i w^j > 0$ if 
	$\, U_i (x) \, w^i = 0\,$,   
\item[(c)]
	there exists a (necessarily unique) vector field $U^i (x)$ such that
	$U^i(x)U_i(x) = -1$ and 
	{\Large $\frac{\partial ^2 \ell(x,v)^2}{\partial v^i v^j}$}
	$U^i (x) = 0\,$.
\end{itemize}
These conditions guarantee that, indeed, the Finsler metric
\begin{equation}\label{gex}
	g_{ij} (x,v) \, := \,  
	\frac{\partial ^2 L(x,v)}{\partial v ^i \partial v ^j} 
	\, = \, \frac{1}{2} \, 
	\frac{\partial^2 \ell (x,v)^2}{\partial v^i \partial v^j} 
	- \, U_i (x) \, U_j (x) 
\end{equation}
has Lorentzian signature, so the Lagrangian (\ref{eq:Lex}) defines a Finsler
spacetime in the sense of Definition \ref{def:finsler}. In this special case,
the set $\big\{ \, (x,v) \in \overset{o}{T} _x X \, \big| \, L(x,v)<0 \, \big\}$ 
has two connected components $Z_x^+M$ and $Z_x^-M$ which are mapped onto each other 
by inversion $(x,v) \mapsto (x, -v)$. If $\ell (x,v)^2$ is a quadratic form 
with respect to $v$, the Finsler metric is independent of $v$. In the more general 
case, the cones $Z_x^+ M$ and $Z_x^-M$ are anisotropic in the sense that we cannot 
find a faithful representation of the rotation group $O(N-1)$, for $\mathrm{dim}(M) 
= N$, by linear transformations of $T_x M$ that leave $Z_x^+ M$ or $Z_x^-M$ invariant. 
-- For this example the Hamiltonian (\ref{eq:defH}) takes the form
\begin{equation}\label{eq:Hex}
	H(x,p) \, = \, \frac{1}{2} \, \big( \, h(x,p)^2 \, - \, 
	U^i (x) \, U^j (x) \, p_i \, p_j \, \big)  
\end{equation}
where $h(x,p)= \ell (x,v)$, with $(x,v) \in \overset{o}{T} _x M$ and $(x,p) \in 
\overset{o}{T}{}^* _x M$ related by (\ref{eq:Legendre}). Hamiltonians of the form
(\ref{eq:Hex}) appear naturally if light propagation in a linear dielectric and 
permeable medium on a general-relativistic spacetime is considered, see \cite{Pe00}, 
eq. (2.73). In general, such a medium is birefringent; there are two types of light 
rays, and each of the two types is governed by a Hamiltonian of the form (\ref{eq:Hex}), 
with the same vector field $U^i (x)$ but different functions $h(x,p)$. 
In this case $U^i (x)$ is the 4-velocity field of the medium and, for each of the two 
types, the function $h(x,p)$ is built in a fairly complicated way from the spacetime 
metric, the dielectricity tensor and the permeability tensor. In this sense, Finsler 
spacetimes with Lagrangians of the form (\ref{eq:Lex}) and, equivalently, 
Hamiltonians of the form (\ref{eq:Hex}) 
have interesting applications in (general-relativistic) optics in media. They can also
be considered as alternative spacetime models, generalizing the formalism of general 
relativity to the (hypothetical) situation that the \emph{vacuum} light rays are 
determined by anisotropic light cones.     		
\end{ex}

\section{Fermat's principle}\label{sec:fermat}

It is now our goal to characterize, in a Finsler spacetime $(M,L)$, the lightlike geodesics
from a point $q$ to a timelike curve $\gamma$ by a variational principle. As the trial 
curves for this variational principle we want to consider all lightlike curves from $q$ to 
$\gamma$. Since, at each point $x \in M$, the Finsler light cone may have arbitrarily many 
components, it will be necessary to restrict to those lightlike curves whose
tangent vectors, on arrival at $\gamma$, belong to the connected component
of the light cone that is selected by the tangent vector of $\gamma$. This
leads to the following definition.

\begin{defin}\label{def:trial}
Choose, in an $N$-dimensional Finsler spacetime $(M,L)$, a point $q \in M$ and a 
$C^{\infty}$ embedding $\gamma : I \rightarrow M$ with $L \big( \gamma (t), 
\dot{\gamma} (t) \big) < 0$, where $I$ is a real interval. For each $t \in I$, 
let $Z_{\gamma (t)}M$ denote the connected component of the set $\big\{ 
\big( \gamma (t),v \big) \in \overset{o}{T} _{\gamma (t)} M \big| L \big( 
\gamma (t) ,v \big) <0 \big\}$ which contains the vector $\big( \gamma (t) , 
\dot{\gamma} (t) \big)$. Define the \emph{space of trial curves} 
${\mathcal{C}}_{q,\gamma}$ as the set of all $C^{\infty}$ maps $\lambda : [0,1] 
\rightarrow M$ with the following properties.
\begin{itemize} 
\item[(a)]
	$\lambda (0) = q \,$.
\item[(b)]
	There is a $\tau ( \lambda ) \in I$ such that $\lambda (1) =
	\gamma \big( \tau (\lambda ) \big)$.
\item[(c)]
	$\lambda$ is lightlike, i.e.
\[
	L \big( \lambda (s) , \dot{\lambda} (s) \big) \, = \, 0 \quad \text{ for all } 
	 \; s \in [0,1] \, ,
\]
	and $\big( \lambda (1) , \dot{\lambda} (1) \big)$ lies in the boundary
	of $Z_{\gamma \left( \tau (\lambda ) \right)}M$. 
\end{itemize}
\end{defin}	

By an \emph{allowed variation} of $\lambda \in {\mathcal{C}}_{q, \gamma}$ we mean a
$C^\infty$ map $\Lambda : \; ] - \varepsilon _0 \, , \, \varepsilon _0 \, [ \; \times
\, [0,1] \, \rightarrow \, M$, $(\varepsilon , s) \mapsto \Lambda (\varepsilon ,s)$
such that $\Lambda (\varepsilon, \cdot ) \in {\mathcal{C}}_{q,\gamma}$ for all $\varepsilon$
and $\Lambda (0 , \cdot )= \lambda$.  

Part (b) of Definition \ref{def:trial} defines the \emph{arrival time
functional} $\tau : {\mathcal{C}}_{q,\gamma} \rightarrow I$. If we have an allowed
variation $\Lambda$ of $\lambda$, we can consider the map $\varepsilon 
\mapsto \tau \big( \Lambda (\varepsilon , \cdot ) \big)$ which maps a real 
interval to a real interval. To link up with the traditional notation of 
variational calculus, in the following we use the symbol $\delta$ for the 
derivative with respect to $\varepsilon$ at $\varepsilon =0 \,$; e.g., we write
\begin{equation}\label{eq:vartau}
	\delta  \tau (\lambda )  \, := \,  
	\frac{d}{d\varepsilon}\Big( \tau \big( \Lambda (\varepsilon , \cdot ) \big) \Big)
	\Big| _{\varepsilon = 0} \; .
\end{equation}

The desired version of Fermat's principle can now be formulated as a mathematical
theorem in the following way.

\begin{theo}\label{theo:fermat}
	{\bf (Fermat's principle for Finsler spacetimes)}
	A curve $\lambda \in {\mathcal{C}}_{q,\gamma}$ is an arbitrarily parametrized
	geodesic if and only if $\delta  \tau (\lambda ) =0$ for all
	allowed variations of $\lambda$ in ${\mathcal{C}}_{q,\gamma}$.
\end{theo}

The proof will be given in the next section.

The statement of Theorem \ref{theo:fermat} can be rephrased in the following way. 
Among all ways to go from $q$ to $\gamma$ at the speed of light, as it is determined 
by the field of light cones selected by $\gamma$ according to part (c) of Definition 
\ref{def:trial}, the light actually chooses those paths that make the arrival time 
stationary. In the isotropic case, i.e., if the Finsler metric is independent of $v$, 
Theorem \ref{theo:fermat} reduces to Kovner's version \cite{Ko90} of Fermat's principle 
which was proven in \cite{Pe90}. In this special case we know from \cite{Pe90} that 
only local minima and saddles, but no local maxima, of the arrival time occur. This 
result is based on the analysis of conjugate points along the respective geodesic. 
One can formulate a Morse index theorem for this situation \cite{Pe95} and, under 
additional assumptions on the global spacetime structure, even set up a full-fledged 
Morse theory \cite{GMP}. It is interesting to investigate whether similar results hold 
in the general, i.e. anisotropic, case of Theorem \ref{theo:fermat}. Such an investigation 
will be postponed to future studies because it requires additional preparatory work on 
the second variational formula. It is true that conjugate points are well-defined and 
their basic properties are well-established whenever one has a Lagrangian with non-degenerate 
Hessian (see, e.g., Morse \cite{Mo73}, Section 1.5), so in particular for the geodesics 
of a Finsler metric with arbitrary signature. However, the finer aspects of the theory,
in particular the relation between the second variation and the number of conjugate
points, have not been worked out for indefinite Finsler metrics so far. (For positive
definite Finsler metrics see Crampin \cite{Cr00, Cr01} and earlier references given therein.)

\section{Proof of Fermat's principle}\label{sec:proof}

We begin with the proof of the `only if' part of Theorem \ref{theo:fermat} which is 
quite easy. So let us assume that $\lambda  \in \mathcal{C} _{q, \gamma}$ is a
geodesic; as $\tau$ is invariant under reparametrizations, we may assume, without
loss of generality, that $\lambda$ is affinely parametrized. As all varied
curves are lightlike, we have for every allowed variation
\begin{equation}\label{eq:varact}
	0 \, = \, \delta \, \int _0 ^1 L \big( \lambda (s) , {\dot{\lambda}} (s)  
	\big) \, ds \, .
\end{equation}
After calculating the $\delta$-differentiation under the integral and integrating 
by parts this leads to
\begin{equation}\label{eq:partint}
	\int _0 ^1 \Big( 
	\, \frac{d}{ds} \, \frac{\partial L \big( \lambda (s), 
	\dot{\lambda} (s) \big)}{ \partial \dot{\lambda}{}^i (s)} \, - \,
	\frac{\partial L \big( \lambda (s) , \dot{\lambda} (s) \big)}{
	\partial \lambda ^i (s)} \, \Big) \, \delta \lambda ^i \, ds \, = \, 
	\frac{\partial L \big( \lambda (s), \dot{\lambda} (s) \big)}{
	\partial \dot{\lambda}{}^i (s)} \, \delta \lambda ^i (s) \Big| _{s=0} ^{\, 1} \; .
\end{equation}
As $\lambda$ is an affinely parametrized geodesic, the bracket under the integral
vanishes. From part (a) and (b) of Definition \ref{def:trial} we find
\begin{gather}
\label{eq:deltalambda0}
	\delta \lambda ^i (0) \, = \, 0 \; , 
\\
\label{eq:deltalambda1}
	\delta \lambda ^i (1) \, = \, \dot{\gamma} {}^i \big( \tau (\lambda ) \big) \,
	\delta \tau ( \lambda ) \; .
\end{gather}
With (\ref{eq:hom2}) this reduces (\ref{eq:partint}) to
\begin{equation}\label{eq:deltatau}
	0 \, = \,  g_{ji} \big(  \lambda (1), \dot{\lambda} (1) \big) \,
	\dot{\lambda} {}^j (s) \, \dot{\gamma} {}^i \big( \tau ( \lambda ) \big) \,
	\delta \tau (\lambda ) \; .
\end{equation}
Part (c) of Proposition \ref{prop:cone} guarantees that $g_{ji} \big(  \lambda (1), 
\dot{\lambda} (1) \big) \, \dot{\lambda} {}^j (s) \, \dot{\gamma}{}^i \big( \tau ( \lambda ) 
\big) \, \neq \, 0 \,$, so we have found that, indeed, $\delta \tau ( \lambda ) = 0$.

The proof of the `if' part of Theorem \ref{theo:fermat} is more involved.
We first establish a lemma that chararacterizes, along a curve $\lambda \in \mathcal{C} 
_{q, \gamma }$, the set of variational vector fields $\delta \lambda (s) = 
\partial _{\varepsilon} \Lambda (\varepsilon , s )|_{\varepsilon = 0}$ that
come from allowed variations $\Lambda$. 

\begin{lem}\label{lem:varfield}
	For $\lambda \in \mathcal{C} _{q, \gamma }$, a $C^{\infty}$ vector field
	$s \mapsto \big( \lambda (s) , A (s) \big)$ along $\lambda$ is the 
	variational vector field of an allowed variation, $A = \delta \lambda$, 
	if and only if 
\begin{gather}
\label{eq:var0}
	A ^i (0) \, = \, 0 \; ,
\\
\label{eq:var1}
	A ^i (1) \, \text{ is a multiple of } \; 
	\dot{\gamma} {}^i \big ( \tau (\lambda ) \big) \; ,
\\
\label{eq:vardiff}
	\frac{\partial L \big( \lambda (s) , \dot{\lambda} (s) \big)}{
	\partial \lambda ^i (s)} \, A ^i (s) \, + \, 
	\frac{\partial L \big( \lambda (s), \dot{\lambda} (s) \big)}{
	\partial \dot{\lambda}{}^i (s)} \, \dot{A}{}^i (s) \, = \, 0 \; .
\end{gather}
\end{lem}
\begin{proof}
Clearly, if $A$ is the variational vector field of an allowed variation,
$A = \delta \lambda$, it has to satisfy the three conditions; this follows 
immediately if we apply the variational derivative $\delta$ to the three 
conditions (a), (b) and (c) of Definition \ref{def:trial}. Now let us assume, 
conversely, that we have a vector field $A$ that satisfies the three conditions. 
It is our goal to construct an allowed variation such that $A = \delta \lambda$. 
We give this construction here only under the additional condition that 
$\lambda$ can be covered by a local coordinate system whose $N$-th basis vector 
field $\partial / \partial x^N$ is timelike, $L \big( x, \partial /  \partial x^N (x) 
\big) < 0$, with $\partial / \partial x^N \big( \gamma ( \tau (\lambda )) \big) = 
\dot{\gamma} \big( \tau ( \lambda ) \big)$. (If this condition is violated, which 
may happen if $\lambda$ has self-intersections, the proof requires to cover $\lambda$ 
with several coordinate patches. The details of this patching procedure, which is 
somewhat awkward although straight-forward, can be carried over from the proof of 
Lemma 2 in \cite{Pe90}.) Using this coordinate system, we construct the desired 
allowed variation $\Lambda$ from the given $A ^i$ in the following way. We define 
the first $(N-1)$ coordinates of $\Lambda$ by
\begin{equation}\label{eq:defLambda}
	\Lambda ^{\alpha} ( \varepsilon , s ) \, = \, \lambda ^{\alpha} (s) \, + \,
	\varepsilon \, A ^{\alpha} (s) \, , \qquad
	\alpha = 1 , \dots , (N-1) \, .
\end{equation}  
With these $(N-1)$ coordinates of $\Lambda$ known, the $N$-th coordinate of 
$\Lambda$ is to be determined by the differential equation
\begin{equation}\label{eq:lightLambda}
	L \big( \Lambda (\varepsilon , s ) , 
	\partial _s \Lambda ( \varepsilon , s) \big) \, = \, 0  
\end{equation}
and the initial condition 
\begin{equation}\label{eq:initialLambda}
	\Lambda ^N( \varepsilon , 0 ) = 0 \, . 
\end{equation}
For $\varepsilon$ sufficiently small, this initial value problem has indeed a 
unique solution $s \mapsto \Lambda ^N (\varepsilon , s)$ on the interval $[0,1]$
which is close to $\lambda^N$. To demonstrate this, we first observe that 
(\ref{eq:lightLambda}) can be locally solved for $\partial _s \Lambda ^N 
(\varepsilon , s )$, because $\frac{\partial L ( x, v)}{\partial v^N} 
\, = \, g_{Nj}(x,v)\, v^j$ is non-zero by part (c) of Proposition \ref{prop:cone} 
for $x = \Lambda ( \varepsilon , s)$ and $v = \partial _s \Lambda (\varepsilon, s)$. 
So the initial value problem has a unique solution on some interval $[0, s_0[ \;$. For 
$\varepsilon =0$, the solution exists up to some $s_0 > 1$ because the curve $\lambda$ 
exists on this interval. By continuity, for all sufficiently small $\varepsilon$ 
the solution exists up to the parameter 1. By construction, all curves $s \mapsto
\Lambda ( \varepsilon, s )$ satisfy the three conditions (a), (b) and (c) of Definition
\ref{def:trial}, so $\Lambda$ is, indeed, an allowed variation of $\lambda$. Finally,
we have to verify that the variational vector field $\delta \lambda$ of this variation 
coincides with the given $A$. This is obvious from (\ref{eq:defLambda}) for the first 
$(N-1)$ coordinates. It is also true for the $N$-th coordinate because both vector fields 
are tangent to the surface $L=0$ which, again by part (c) of Proposition \ref{prop:cone}, 
is transverse to the $N$-lines of our coordinate system.
\end{proof}

We are now ready to prove the `if' part of Theorem \ref{theo:fermat}.
So assume that at $\lambda \in \mathcal{C} _{q , \gamma }$ the condition
$\delta \tau (\lambda ) = 0$ holds for all allowed variations. Let $s \mapsto 
\big( \lambda (s) , B (s) \big)$ be any $C^{\infty}$ vector field
along $\lambda$ that vanishes at the endpoints, $B(0)=0$ and $B(1)=0$. 
We choose a vector field $s \mapsto \big( \lambda (s) , U (s) \big)$ along 
$\lambda$ with $L \big( \lambda (s) , U (s) \big) < 0$ for all $s \in [0,1]$ and
$U^i (1) = \dot{\gamma}{}^i \big( \tau ( \lambda ) \big)$. (Such a vector
field exists because $L$ takes negative values on an open set.) We define a
function $f:[0,1] \rightarrow \mathbb{R}$ by the differential equation
\begin{gather}\label{eq:deff}
	\frac{\partial L \big( \lambda (s) , \dot{\lambda} (s) \big)}{
	\partial \lambda ^i (s)} \, \Big( \, B^i (s) \, + \, 
	f(s) \, U^i (s) \, \Big) \, + \, 
\\
\nonumber
	\frac{\partial L 
	\big( \lambda (s), \dot{\lambda} (s) \big)}{
	\partial \dot{\lambda}{}^i (s)} \, \Big( \, \dot{B}{}^i (s) \, 
	+ \, f(s) \, \dot{U}{}^i (s) \, + \, \dot{f} (s) \, U^i (s) 
	\, \Big) = \, 0 
\end{gather}
and the initial condition $f(0)=0$. Part (c) of Proposition \ref{prop:cone} 
guarantees that 
\begin{equation}\label{eq:solvedotf}
\frac{\partial L \big( \lambda (s), \dot{\lambda} (s) \big)}{
\partial \dot{\lambda}{}^i (s)} \, U^i (s) = g_{ji} \big( \lambda (s), 
\dot{\lambda} (s) \big) \dot{\lambda}{}^j (s) U^i(s) \neq 0 \, ,
\end{equation}
so (\ref{eq:deff}) can be solved for $\dot{f}(s)$ and the initial value problem 
has, indeed, a unique solution. If we now define 
\begin{equation}\label{eq:varB}
	A ^i (s) \, := \, B^i (s) \, + \, f(s) \, U^i (s)
\end{equation}
we immediately verify from Lemma \ref{lem:varfield} that it comes from an allowed
variation, so we may write $A^i = \delta \lambda ^i$. For this variation we find
$\delta \tau ( \lambda ) = f(1)$ from condition (b) of Definition \ref{def:trial}. 
Next we define a function $h: [0,1] \rightarrow \mathbb{R}$ by the differential 
equation
\begin{gather}\label{eq:defh}
	\frac{\partial L 	\big( \lambda (s), \dot{\lambda} (s) \big)}{
	\partial \dot{\lambda}{}^i (s)} \, U^i (s) \, \dot{h} (s) \, = 
\\
\nonumber 
	\frac{\partial L \big( \lambda (s), \dot{\lambda} (s) \big)}{
	\partial \dot{\lambda}{}^i (s)} \, \dot{U}{}^i (s) \, + \, 
	\frac{\partial L \big( \lambda (s) , \dot{\lambda} (s) \big)}{
	\partial \lambda ^i (s)} \, U^i (s) \, - \, \frac{d}{ds}
	\Big( \frac{\partial L 	\big( \lambda (s), \dot{\lambda} (s) \big)}{
	\partial \dot{\lambda}{}^i (s)} \, U^i (s) \, \Big)
\end{gather}
and the initial condition $h(0)=0$. As above for the function $f$, part (c) of 
Proposition \ref{prop:cone} guarantees that this initial value problem has a 
unique solution. By multiplying (\ref{eq:deff}) with the integrating factor 
$e^{h(s)}$ we get
\begin{gather}\label{eq:hfac}
	e^{h(s)} \, \Big( 
	\frac{d}{ds} \, 
	\frac{\partial L \big( \lambda (s), \dot{\lambda} (s) \big)}{
	\partial \dot{\lambda}{}^i (s)} \, - \,	
	\frac{\partial L \big( \lambda (s) , \dot{\lambda} \big)}{
	\partial \lambda ^i (s)} \, + \, 
	\dot{h}(s) \, 
	\frac{\partial L \big( \lambda (s), \dot{\lambda} (s) \big)}{
	\partial \dot{\lambda}{}^i} \, \Big) \, B^i (s) \, = 
\\
\nonumber 
	\frac{d}{ds} \, \Big( \, e^{h(s)} \, \frac{\partial L \big( \lambda (s), 
	\dot{\lambda} (s) \big)}{\partial \dot{\lambda}{}^i (s)} \,
	\big( \, B^i (s) \, + \, f(s) \, U^i (s) \, \big) \Big) \, .
\end{gather} 
Integration of this equation from 0 to 1 yields, owing to the
boundary conditions $B^i(0)=0$, $B^i(1)=0$, $f(0)=0$, and $f(1)=
\delta \tau ( \lambda )$:
\begin{gather}\label{eq:hint}
	\int _0 ^1 {e^{h(s)} \, \Big( 	
	\frac{d}{ds} \, 
	\frac{\partial L \big( \lambda (s), \dot{\lambda} (s) \big)}{
	\partial \dot{\lambda}{}^i (s)} \, - \, 
	\frac{\partial L \big( \lambda (s) , \dot{\lambda} \big)}{
	\partial \lambda ^i (s)} \, + \, 
	\dot{h}(s) \, 
	\frac{\partial L \big( \lambda (s), \dot{\lambda} (s) \big)}{
	\partial \dot{\lambda}{}^i} \, \Big) \, B^i (s) \, ds } \,  = 
\\
\nonumber 
	e^{h(1)} \, \frac{\partial L \big( \lambda (1), 
	\dot{\lambda} (1) \big)}{\partial \dot{\lambda}{}^i (1)} \,
	 \, U^i (1) \, \delta \tau ( \lambda )  \, .
\end{gather} 
By hypothesis, $\delta \tau ( \lambda) = 0$; so we have demonstrated that
the left-hand side of (\ref{eq:hint}) is zero for any $B^i$ that vanishes
at both end-points. Hence, the fundamental lemma of variational calculus
implies that the bracket under the integral is zero, i.e., that $\lambda$
satisfies (\ref{eq:ELpar}) which is the defining equation for an arbitrarily 
parametrized geodesic.

\section{Some special cases}\label{sec:stat}
A major difference of our variational principle, in comparison to standard 
variational principles with relevance to physics, is in the fact that the
functional to be varied, that is the arrival time, is not given as an 
integral over the trial curves. In this section we will specialize to situations
where it is indeed possible to rewrite the arrival time as such an integral.
This will bring our variational principle closer to the standard literature
on variational calculus and, at the same time, it will clarify the relation
of our variational principle to some earlier versions of Fermat's principle.   

To that end we specialize to the case that our $N$-dimensional Finsler spacetime 
$(M,L)$ can be covered by a single chart in which the $N$th coordinate vector field 
$\partial / \partial x^N$ is timelike, $L \big( x, \partial / \partial x^N (x) \big)$. 
(Actually, it would be sufficient for the following reasoning to assume that $M$ is 
an open subset of a fiber bundle, with timelike fibers diffeomorphic to $\mathbb{R}$. 
However, for notational convenience we will restrict to the more special case.) We 
can then consider our variational principle for the case that $q$ is an arbitrary 
event in $M$ and $\gamma$ is an integral curve of $\partial / \partial x^N$, i.e.
$\dot{\gamma} = \partial / \partial x^N \circ \gamma$. Now along each trial curve 
$\lambda$ for our varitional principle the equation
\begin{equation}\label{eq:L0}
  L\big(  \boldsymbol{\lambda} (s) ,  t(s) ,
  \dot{\boldsymbol{\lambda}} (s) ,  dt(s)/ds \big)
  \, = \, 0
\end{equation}
holds, where we have written
\begin{equation}\label{eq:bold}
  \boldsymbol{\lambda} (s) \, = \, \big( \lambda ^1 (s) , \dots , 
  \lambda ^{N-1} (s) \big) \: , \qquad t(s) \, = \, \lambda ^N (s) \, \; .
\end{equation}
Proposition  \ref{prop:cone} (c) guarantees that (\ref{eq:L0}) can be 
solved for $dt(s)/ds$ along every trial curve,
\begin{equation}\label{eq:solvef}
  \frac{dt(s)}{ds} \, = f \big( \boldsymbol{\lambda} (s) , t(s) , 
  \dot{\boldsymbol{\lambda}} (s) \big) \, ,
\end{equation}
which defines a function $f$. Now, owing to the fact that $\gamma$ is an 
integral curve of $\partial / \partial x^N$, the arrival time is the 
same as the travel time measured in terms of the coordinate $x^N$, up
to a number that is the same for all trial curves,
\begin{equation}\label{eq:taut}
	\tau ( \lambda ) \, = \, \int _0 ^1 \frac{dt(s)}{ds} ds \,
	+ \, {\mathrm{constant}} \, . 
\end{equation}
Hence, by (\ref{eq:solvef}) our variational principle takes the form
\begin{equation}\label{eq:tauspace}
	\delta \,  \int _0 ^1 \, f \big( \boldsymbol{\lambda} (s) , t(s) , 
  	\dot{\boldsymbol{\lambda}} (s) \big) \, ds \, = \, 0 \, .
\end{equation}
This is precisely the variational principle of Godin and Voronovich
\cite{GV04}. (In the case of an isotropic Finsler spacetime it reduces
to the variational principle given in Theorem 3 of \cite{Pe90}.) We 
have thus shown that this variational principle of Godin and Voronovich
is a special case of our version of Fermat's principle, formulated in 
Theorem \ref{theo:fermat}. This special case differs in two respects 
from the more general version of Theorem \ref{theo:fermat}. First, the 
variational functional is now written as an integral. Second, the trial curves 
are now curves $\boldsymbol{\lambda}$ in space, rather than in spacetime; each 
curve $\boldsymbol{\lambda}$ starts at the spatial point to which the spacetime 
point $q$ projects and terminates at the spatial point to which the spacetime 
curve $\gamma$ projects. So it is a purely spatial variational principle
for curves between two fixed points. However, in the integrand of (\ref{eq:tauspace}) 
the function $t(s)$ appears. This function has to be determined, for each trial 
curve $\boldsymbol{\lambda}$, by solving the differential equation (\ref{eq:solvef}) 
with the initial condition $t(0) = t_0$, where $t_0$ is the $x^N$ coordinate of 
the spacetime point $q$. Only after $t(s)$ has been determined for each trial curve 
can the variational principle (\ref{eq:tauspace}) be set into action. (Trial curves
for which $t(s)$ is not defined on the whole interval $[0,1]$ have to be discarded.)
In most cases, determining $t(s)$ is very awkward if not impossible; so it is usually 
recommendable to stick with the more general spacetime version of our variational 
principle, as given in Theorem \ref{theo:fermat}, rather than to switch to the more 
special spatial version of (\ref{eq:tauspace}), even in cases where the latter holds 
true.

There is one situation, however, in which the spatial version is indeed much
more convenient, namely if there is a function $e^{h(x,v)}$ such that
\begin{equation}\label{eq:Killing}
	\frac{\partial}{\partial x^N} \, \big( e^{2h(x,v)} L(x,v) \big) \,
	= \, 0 \, .
\end{equation}
In this case, we call $\partial / \partial x^N$ a \emph{generalized conformal
Killing vector field}. (If (\ref{eq:Killing}) holds with a function $h$ that
is independent of $v$,  $\partial / \partial x^N$ is called a \emph{conformal
Killing vector field}, and if (\ref{eq:Killing}) holds with $h$ identically
equal to zero, $\partial / \partial x^N$ is called a \emph{Killing vector field}.)
Then the function $f$ of (\ref{eq:solvef}) is independent of $t(s)$, i.e., 
the variational principle (\ref{eq:tauspace}) takes the form
\begin{equation}\label{eq:taustat}
	\delta \,  \int _0 ^1 \, f \big( \boldsymbol{\lambda} (s) , 
  	\dot{\boldsymbol{\lambda}} (s) \big) \, ds \, = \, 0 \, .
\end{equation}
This is a purely spatial variational principle that does not involve the
necessity to solve additional differential equations. It is easy to verify
that the homogeneity of $L$ implies that $f$ is positively homogeneous of 
degree one, 
\begin{equation}\label{eq:homf}
  	f \big( \boldsymbol{x} , k \boldsymbol{v} \big) \, = \, 
  	k \, f \big( \boldsymbol{x} , \boldsymbol{v} \big) \quad 
	\mathrm{for} \: k  \in \; ]0, \infty [ \: ,
\end{equation} 
so the functional in (\ref{eq:taustat}) is invariant under reparametrization. 
If we add the assumption that the Hessian of $f$ with respect to the (purely 
spatial) velocity coordinates is positive definite, (\ref{eq:taustat}) is 
equivalent to varying the length functional of a positive definite Finsler 
metric; then the solution curves ${\boldsymbol{\lambda}}$ are, of course, the 
geodesics of this positive definite Finsler metric. The variational principle 
(\ref{eq:taustat}) is of the same form as the time-independent versions of 
Fermat's principle that have been discussed, for light rays in anisotropic 
media, in \cite{KK65, Ne83, Gl80} and, for sound rays in anisotropic media, 
in \cite{Ba61, Ug72, ES92, Ce02}. 

This construction also works the other way round. We can start with a
function $f ( \boldsymbol{x} , \boldsymbol{v} )$ that satisfies
the homogeneity condition (\ref{eq:homf}) and has a positive definite
Hessian with respect to the velocity coordinates. We can then  define
a spacetime Lagrangian
\begin{equation}\label{eq:defL}
	L ( \boldsymbol{x}, x^N , \boldsymbol{v}, v^N ) \, = \, \frac{1}{2}
	\big( \, 	f ( \boldsymbol{x}, \boldsymbol{v} )^2 \, - \, (v^N)^2 \, 
	\big) \; . 
\end{equation}
$L$ gives us a Finsler spacetime for which $\partial / \partial x^N$ is 
a Killing vector field. (Note that this is a special case of the Lagrangian
considered in Example \ref{ex:hamilton}.)	Our version of Fermat's principle
says that the lightlike geodesics of this Finsler spacetime project to the
geodesics of the positive definite spatial Finsler structure given by $f$. 
Thus, our variational principle encompasses, in a spacetime formulation, 
all time-independent versions of Fermat's principle where the spatial rays 
are the geodesics of a positive definite spatial Finsler structure.  

\section{Outlook}\label{sec:out}
The Fermat principle in Finsler spacetimes presented in this paper is a 
satisfactory formulation for rays in time-dependent anisotropic situations, 
as long as dispersion does not occur. It conveniently comprises many earlier 
versions in a geometrical spacetime setting. However, some questions
are still open.

As already mentioned at the end of Section \ref{sec:fermat}, the second 
variation formula for our variational principle has not been evaluated so
far. This is of relevance to the question  of whether a solution curve
is a local minimum, a local maximum or a saddle of the arrival time functional.
It would be desirable to investigate whether the index of the second 
variation is related to the number of conjugate points, in analogy to the
Morse index theorem of the pseudo-Riemannian case.   

There are two more technical generalizations of our Fermat principle 
which have not been worked out so far. First, in this paper we have 
restricted to Finsler spacetimes of class $C^{\infty}$, and we have 
formulated Fermat's principle for trial curves of class $C^{\infty}$.
For some applications it might be recommendable to consider piecewise smooth
Finsler structures and piecewise smooth (``zig-zag'') trial curves. We have
not done this here because it makes the proof considerably more cumbersome. 
Second, it is likely that the non-degeneracy of the Finsler metric could 
be a little bit relaxed. E.g., the Lagrangian 
\begin{gather}\label{eq:degex}
L: \overset{o}{T} {\mathbb{R}}{}^N \; \rightarrow \; {\mathbb{R}}, 
\\
\nonumber
(x,v) \: \mapsto \: 
\sqrt{\, (v^1)^4 + \dots + (v^{N-1})^4 \, } \, - \, (v^N)^2 
\end{gather}
violates condition (b)(ii) of Definition \ref{def:finsler} because the
Finsler metric degenerates on the $v^{\mu}$-axis, for each $\mu = 1, \dots, N-1$;
hence, this case is not within the class of Lagrangians for which we have proven
Fermat's principle in this paper. However, as the set of lightlike vectors for 
which the non-degeneracy condition is violated is a set of measure zero, it might 
be possible to show by a continuity argument that Fermat's principle is still 
valid in this case and in similar cases. 

Finally, it should be stressed again that the formulation of ray propagation
in terms of Finsler geometry excludes dispersion, i.e., it does not apply
to cases where the propagation of rays depends on frequency. If rays are derived 
from a Hamiltonian on the cotangent bundle over spacetime, dispersion is 
absent whenever the Hamiltonian is positively homogeneous (see, e.g., 
\cite{Pe00}, p.116), as is inherent in Finsler geometry. Therefore, any 
formulation of ray propagation that includes dispersion has to leave the
domain of Finsler geometry. It was already mentioned that a version of
Fermat's principle allowing for time-dependence, anisotropy and dispersion
was brought forward in \cite{Pe00}, Section 7.3 (and that another such 
version was suggested by Voronovich and Godin \cite{VG03}). The problem 
with this version is that it is a variational principle for curves in the 
cotangent bundle over spacetime, not for curves in spacetime. The condition
under which it can be reduced to a variational principle for curves in 
spacetime is given in \cite{Pe00}, Section 7.3. However, this is only a 
statement on existence; even if one has verified, for a particular case, 
that this condition holds true, it is not obvious how to get an explicit 
formulation of the variational principle in terms of curves in spacetime. 
E.g., such an explicit formulation was worked out for rays in a non-magnetized 
plasma, which is an example of a dispersive and isotropic medium, in \cite{Pe01}.
On the other hand, for a magnetized plasma, which is an example of a dispersive 
and anisotropic medium, such a formulation does not exist so far. In view of
applications to astrophysics, this is the most interesting case for which a 
spacetime formulation of Fermat's principle in the spirit of Kovner, allowing 
for arbitrary time-dependence, is still to be worked out. 


\end{document}